\documentclass{article}
\usepackage{spconf,amsmath,graphicx,url}
\usepackage{booktabs, multirow}
\usepackage{subcaption}
\usepackage[table]{xcolor}
\usepackage{amssymb}
\usepackage{url}

\title{Stethoscope-guided Supervised Contrastive Learning for Cross-domain Adaptation on Respiratory Sound Classification}

\name{June-Woo Kim$^{1,5}$\thanks{\hspace{-1.7em}$^{\dagger}$corresponding author, This research was supported by the MSIT(Ministry of Science and ICT), Korea, under the ITRC(Information Technology Research Center) support program(IITP-2024-2020-0-01808) supervised by the IITP(Institute of Information \& Communications Technology Planning \& Evaluation), and by Brian Impact, a non-profit organization dedicated to the advancement of science and technology.}, Sangmin Bae$^{2,5}$, Won-Yang Cho$^{3,5}$, Byungjo Lee$^{4,5}$, Ho-Young Jung$^{1\dagger}$
}

\address{Author Affiliation(s)}

\address{
  $^1$Department of Artificial Intelligence, Kyungpook National University \\ 
  $^2$KAIST AI\quad 
  $^3$SMARTSOUND\quad
  $^4$National Cancer Center\quad
  $^5$RSC LAB, MODULABS\\
  \{kaen2891, hoyjung\}@knu.ac.kr
  }

\usepackage[colorlinks=true,linkcolor=cyan,urlcolor=red]{hyperref}
\hypersetup{
    colorlinks=true,
    linkcolor=black,
    filecolor=magenta,
    citecolor=black,
    urlcolor=black,
    }
\begin{document}
%
\maketitle
\begin{abstract} 
Despite the remarkable advances in deep learning technology, achieving satisfactory performance in lung sound classification remains a challenge due to the scarcity of available data. Moreover, the respiratory sound samples are collected from a variety of electronic stethoscopes, which could potentially introduce biases into the trained models. When a significant distribution shift occurs within the test dataset or in a practical scenario, it can substantially decrease the performance. To tackle this issue, we introduce cross-domain adaptation techniques, which transfer the knowledge from a source domain to a distinct target domain. In particular, by considering different stethoscope types as individual domains, we propose a novel stethoscope-guided supervised contrastive learning approach. This method can mitigate any domain-related disparities and thus enables the model to distinguish respiratory sounds of the recording variation of the stethoscope. The experimental results on the ICBHI dataset demonstrate that the proposed methods are effective in reducing the domain dependency and achieving the ICBHI Score of 61.71\%, which is a significant improvement of 2.16\% over the baseline.  
\end{abstract}

\section{Introduction}
\label{sec:intro}

An {\it electronic stethoscope} is a device utilized by medical professionals to analyze sounds within the human body, including those produced by the heart, lungs, and other organs.
In general, the design and construction of the electronic stethoscope can influence its efficacy in capturing and recording respiratory sounds~\cite{rennoll2020electronic, seah2023review}.
Despite the significant advancements originating from artificial intelligence and signal-processing domain in respiratory sound analysis~\cite{gairola2021respirenet, ren2022prototype, wang2022domain, nguyen2022lung, bae23b_interspeech}, the type and quality of the stethoscope used, as well as the characteristics of the surrounding environment, can introduce bias into the recorded breath sounds and affect their interpretation.
This may lead to a shift in data distribution~\cite{guan2021domain}, resulting in biased representations within neural network models and hindering their ability to generalize~\cite{quinonero2008dataset, torralba2011unbiased}.

\begin{figure}[!t]
    \vspace{-3mm}
    \centering
    \includegraphics[width=0.92\linewidth]{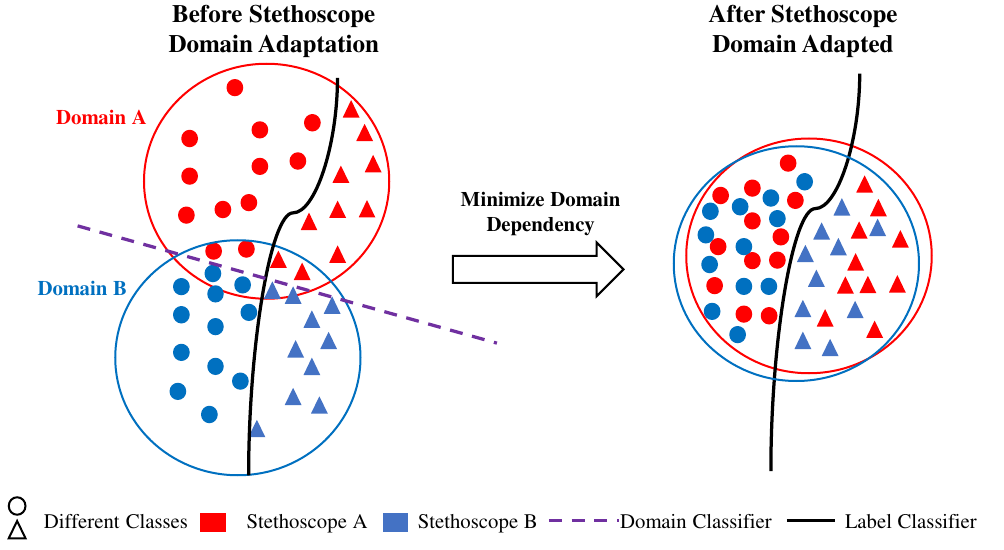}
    \caption{Overview of the proposed stethoscope-guided cross-domain adaptation. The proposed method attempts to minimize the dependency or distribution shift between respiratory sounds recorded by different stethoscopes.}
    \label{fig:overview}
    \vspace{-5mm}
\end{figure}

\begin{figure*}[t!]
    \vspace{-5mm}
    \centering
    \includegraphics[width=1.0\linewidth]{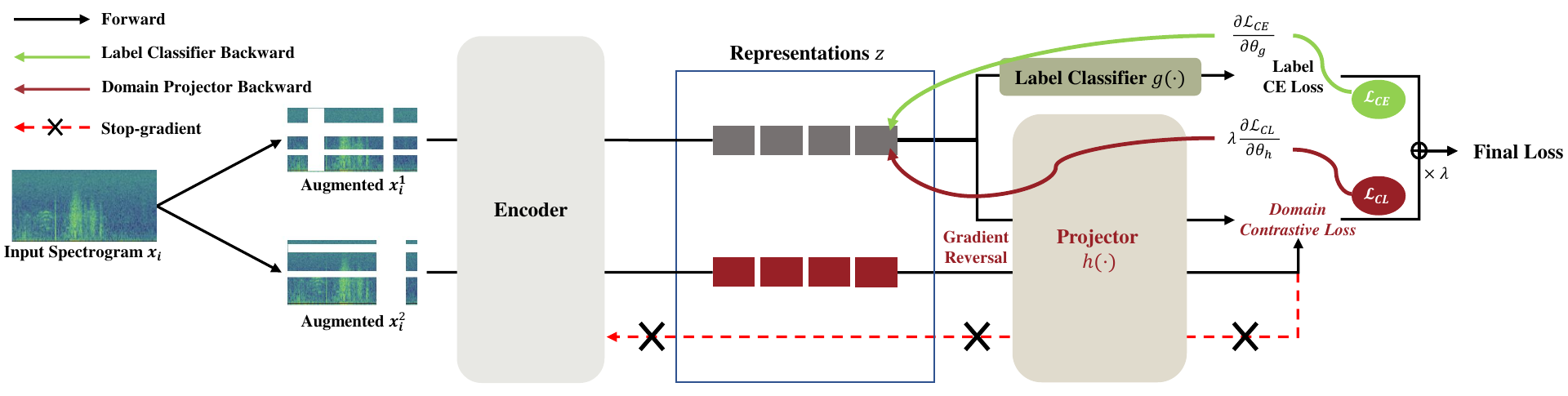}
    \vspace{-3mm}
    \caption{Overall illustration of proposed stethoscope-guided supervised contrastive learning for cross-domain adaptation.}
    \label{fig:proposed}
    \vspace{-5mm}
\end{figure*}

To address this, we focus on the \textit{domain adaptation} method that uses data collected from the source domain to train the model for the target domain, as in Fig.~\ref{fig:overview}. The domain adaptation techniques facilitate the training of effective models capable of extracting domain-invariant features, thereby overcoming the challenges posed by domain dependencies in the data distribution. Therefore, in the context of respiratory sound classification, we consider each type of stethoscope as an individual domain for cross-domain adaptation. Although there are differences in the perception of respiratory sounds across different stethoscopes, we aim to consistently learn the characteristics of both abnormal and normal sounds, even in the presence of mixed samples in data distribution.

In this paper, we introduce two methods for cross-domain adaptation in a supervised manner. 
First, we present a straightforward methodology that leverages domain adversarial training inspired by~\cite{ganin2016domain}. 
And, for more effective cross-domain adaptation, we propose a novel approach that simultaneously considers both domain adversarial training and stethoscope-guided supervised contrastive learning~\cite{khosla2020supervised}. 
As shown in Fig.~\ref{fig:proposed}, the extracted representations of a multi-viewed batch~\cite{khosla2020supervised} are computed using supervised contrastive loss, and their gradients are multiplied by a negative constant during the back-propagation process to reverse the gradients.
The gradient reversal feature extractor forces the model to reduce the distribution shift between different stethoscope classes while maintaining equivalence in the same class.

Experimental results demonstrate that the proposed method achieves the ICBHI~\cite{rocha2018alpha} Score of 61.71\% which represented 2.16\% improvement over the baseline and exhibited performance comparable to the state-of-the-art on the ICBHI dataset. 
This suggests that the proposed method can be an effective approach for addressing domain dependencies in terms of domain adaptation\footnote{Code is available at~\url{https://github.com/kaen2891/stethoscope-guided_supervised_contrastive_learning}}.

\vspace{-5mm}
\section{Preliminaries}

\label{sec:Pre}
\subsection{Details of ICBHI Dataset}
As shown in Table~\ref{tab:data_datails}, the ICBHI~\cite{rocha2018alpha} dataset was constructed using four distinct stethoscope types: \textit{Meditron, LittC2SE, Litt3200}, and \textit{AKGC417L}, each with its own unique input sensor.
The Litt3200 uses a piezoelectric sensor to capture low-frequencies like heart murmurs, while the AKGC417L and Meditron use condenser microphones which are suitable for high-frequencies like wheezes, with superior resolution for wider frequencies. The LittC2SE is an analog stethoscope with a dynamic microphone ideal for mid-frequencies to strike a balance for general auscultation.

In lung sound labels, there are four classes in train/test sets: \textit{normal, crackle, wheeze,} and \textit{both} (crackle + wheeze). In stethoscope labels, however, the absence of LittC2SE data in the test set as well as the dependency on data quantity between the training and test sets suggests that conventional training methods might be biased, potentially leading to performance degradation.



\begin{table}[!t]
    \centering
    \caption{Overall details of the ICBHI lung sound dataset. We report two perspectives: lung sound and device (stethoscope).}
    \label{tab:data_datails}
    \addtolength{\tabcolsep}{10pt}
    \resizebox{\linewidth}{!}{
    \begin{tabular}{clccc}
    \toprule
    & label & train & test & sum \\
    \hline 
    \multirow{4}{*}{\multirow{4}{*}\textbf{lung}} & Normal & 2,063 & 1,579 & 3,642 \\
    & Crackle & 1,215 & 649 & 1,864 \\
    & Wheeze & 501 & 385 & 886 \\
    & Both & 363 & 143 & 506 \\
    \bottomrule
    
    \multirow{4}{*}{\multirow{4}{*}\textbf{device}} & Meditron & 997 & 459 & 1,459 \\
    & LittC2SE & 594 & 0 & 594 \\
    & Litt3200 & 41 & 461 & 502 \\
    & AKGC417L & 2,510 & 1,836 & 4,346 \\
    \bottomrule
    \end{tabular}}
    \vspace{-5mm}
\end{table}

\vspace{-3mm}
\subsection{Preprocessing Details}

We followed the pretrained Audio Spectrogram Transformer~\cite{gong21b_interspeech} (AST) settings for the ICBHI dataset with SpecAugment~\cite{park2019specaugment} as presented in~\cite{bae23b_interspeech}. Specifically, we resampled all lung sound samples to 16 kHz, fixed their lengths to 8 seconds, and extracted 128-dim Fbank features with window/overlap sizes of 25/10ms, respectively. We also normalized the features using the mean and standard deviation of --4.27 and 4.57~\cite{gong21b_interspeech}.

\vspace{-3mm}
\subsection{Training Details}\label{sec:training_details}
We used the official split of the ICBHI dataset (train-test split as 60-40\%), and the default experimental setup for the AST and other architectures as described in~\cite{bae23b_interspeech}. We utilized a weighted cross-entropy (CE) loss~\cite{gairola2021respirenet, bae23b_interspeech, moummad2022supervised} to alleviate the imbalanced data classification problem, except when employing the proposed domain adaptation methods. 
For the domain classifier in domain adversarial training, we used a 4-class linear head. 
For stethoscope-guided supervised contrastive learning, we utilized a projector that consists of two 768-dimensional MLP layers, followed by ReLU and BN layers. 
We set the temperature parameter $\tau=0.06$, and
the domain adaptation parameter $\lambda$ is initiated at 0.0096 and is gradually increased to 1.0 until reaching 200 training steps with a batch size of 8 and 50 epoch settings.
We reported results using a fixed set of five seeds for our experiments.


\vspace{-3mm}
\subsection{Evalutation Metrics}
We used \textit{Sensitivity} ($S_{e}$), \textit{Specificity} $(S_{p})$, and \textit{Score} as the performance metrics for the ICBHI dataset~\cite{rocha2018alpha}. Sensitivity, also known as the true positive rate, indicates the percentage of actual disease cases that are accurately detected. Specificity, on the other hand, indicates the percentage of healthy individuals who are rightly classified as disease-free. We used the Score metric that is calculated by the arithmetic mean of Specificity and Sensitivity to achieve a balance between them.

\vspace{-3mm}
\section{Proposed Method}
\label{sec:proposed}
\vspace{-1mm}
\subsection{Baselines}
The AST~\cite{gong21b_interspeech} has shown superior performance on a variety of audio-related tasks, owing to its pretrained weights derived from large-scale datasets such as ImageNet~\cite{deng2009imagenet} and AudioSet~\cite{audioset}. AST has also been used in studies of lung sound classification, achieving state-of-the-art performance using a simple fine-tuning approach~\cite{bae23b_interspeech}. 
Therefore, we use the AST model setting described in~\cite{bae23b_interspeech} as a baseline for our respiratory sound classification task.

\vspace{-3mm}
\subsection{Domain Adversarial Training}
\vspace{-1mm}
To address the variations in data distribution between different stethoscope domains, we propose a simple stethoscope Domain Adversarial Training (DAT) approach motivated by~\cite{ganin2016domain}. The proposed DAT is comprised of two losses:
\vspace{-1mm}
\begin{align}\label{eq:sdat1} 
\mathcal{L}_{\text{CE}} \! = -\sum_{i=1}^n\! \, y_{i}\! \, \log \, \!(\hat{y_{i}}), \quad \mathcal{L}_{\text{DA}} \! = -\sum_{i=1}^n\! \, d_{i}\! \, \log \, \!(\hat{d_{i}}).
\end{align}
where $\mathcal{L}_{\text{CE}}$ and $\mathcal{L}_{\text{DA}}$ are CE loss with label $y$ and stethoscope domain label $d$ (division by $N$ is omitted), and the predicted probabilities $\hat{y}$ and $\hat{d}$ are obtained by passing through the class and domain classifiers, respectively.
We formulate the proposed DAT as: $\mathcal{L}_{\text{DAT}} = \mathcal{L}_{\text{CE}} + \lambda \, \! \mathcal{L}_{\text{DA}}$ where $\lambda$ is a domain regularization parameter drawn from~\cite{ganin2016domain}. In other words, the goal of DAT is to minimize the classification error for the respiratory sound class and to ensure that the learned features cannot distinguish between the stethoscope domains. 

\vspace{-1mm}
\subsection{Proposed Method: SG-SCL}\label{sec:SG-SCL}
For more effective learning, we propose a novel approach that simultaneously considers both domain adversarial training and \textbf{S}tethoscope-\textbf{G}uided \textbf{S}upervised \textbf{C}ontrastive \textbf{L}earning (SG-SCL) for cross-domain adaptation. In detail, the loss for the proposed SG-SCL can be formulated as $\mathcal{L}_{\text{CE}}\,+\,\!\, \lambda\!\, \mathcal{L}_{\text{CL}}$, where $\mathcal{L}_{\text{CL}}$ is as follows:
\vspace{-1mm}
\begin{align}\label{eq:sg-scl1}
\sum_{i \in I} - \log \Big\{\!\frac{1}{|P(i)|\!}\sum_{p \in P(i)}\! \frac{e \, \! (h(z_{i}) \! \cdot \texttt{sgd}\, (h(z_{p})) \! \, / \tau)}{\sum_{a \in A(i)} \! e \, \!(h(z_{i}) \cdot \texttt{sgd}\, (h(z_{a})) \! \, /\tau)} \Big\} 
\end{align}
where the index $i$ is the anchor index from $A(i) \equiv \, I \setminus \{i\}$, and $P(i) \equiv \, \{p\in \! A(i) :d_{p} = d_{i} \}$ represents the collection of all positive samples within the multi-viewed batch that corresponds to the $i$-th sample~\cite{khosla2020supervised}. $e$ and $\texttt{sgd}$ denote the exponential function and stop-gradient\footnote{is a technique employed to selectively prevent the backpropagation algorithm from updating the weights of certain layers during training.} operation, respectively.
$h$ is the projector described in Section~\ref{sec:training_details}, which takes the feature extractor outputs $z$ as input. Both $z$ and $h$ have the same dimension, and all $z$ are normalized before the dot product.
The $\tau$ controls the sharpness of the cosine similarity.


In other words, the augmented samples from the multi-viewed batch are first fed into the encoder and then computed with Eq.~\ref{eq:sg-scl1}. 
This encourages the model to reduce the dependencies between distinct classes while maintaining equivalence in the same class. Note that the \textit{gradient reversal} in Fig.~\ref{fig:proposed} depicted that the gradients are multiplied by a negative constant during the back-propagation process.

\begin{table}[!t]
    \centering
    \vspace{-1mm}
    \caption{SG-SCL performance based on two factors: anchor representation $z_{i}$ and that of target $z_{p}$. $\footnotesize \texttt{sgd}$ indicate the stop-gradient operation. \textbf{Bold} denotes the best result. The final implementation choices were highlighted. 
    }\label{tab:target_representations}
    \renewcommand{\arraystretch}{1.2}
    \addtolength{\tabcolsep}{0pt}
    \scriptsize{
    \begin{tabular}{lllllll}
    \toprule
    anchor & target & $S_p$\,(\%) & $S_e$\,(\%) & \textbf{Score}\,(\%) \\
    \midrule

    $z_{i}$ & $z_{p}$ & $\text{\bf 89.84}_{\pm 3.92}$ & $\text{{13.61}}_{\pm 5.67}$ & $\text{51.73}_{\pm 1.53}$ \\ 
    $z_{i}$ & $h(z_{p})$ & $\text{76.30}_{\pm 1.55}$ & $\text{\bf 44.60}_{\pm 2.20}$ & $\text{60.45}_{\pm 0.44}$ \\ 
    $h(z_{i})$ & $z_{p}$ & $\text{81.87}_{\pm 3.20}$ & $\text{39.83}_{\pm 1.05}$ & $\text{60.85}_{\pm 1.60}$ \\ 
    $h(z_{i})$ & $h(z_{p})$ & $\text{{77.25}}_{\pm 3.43}$ & $\text{{36.35}}_{\pm 17.97}$ & $\text{{60.78}}_{\pm 0.85}$ \\
    $h(z_{i})$ & \texttt{sgd}$(z_{p})$ & $\text{{76.31}}_{\pm 6.35}$ & $\text{{43.79}}_{\pm 4.38}$ & $\text{60.05}_{\pm 1.19}$ \\ 
    \rowcolor[gray]{0.85}    
    $h(z_{i})$ & \texttt{sgd}$(h(z_{p}))$ & $\text{79.87}_{\pm 8.89}$ & $\text{{43.55}}_{\pm 5.93}$ & $\text{\bf 61.71}_{\pm 1.61}$ \\ 









    \bottomrule
    \end{tabular}
    }
    
    \vspace{-3mm}
\end{table}
\begin{table}[!t]
    \centering
    \caption{Respiratory sound classification performance of different architectures using CE, DAT, and SG-SCL. Here, IN and AS refer to ImageNet~\cite{deng2009imagenet} and AudioSet~\cite{audioset}, respectively. \textbf{Bold} denotes the best result.}\label{tab:comparison1}
    \vspace{-1mm}
    \addtolength{\tabcolsep}{0pt}
    \resizebox{\linewidth}{!}{
    \begin{tabular}{llclll}
    \toprule
    architecture & method & pretrain & $S_p$\,(\%) & $S_e$\,(\%) & \textbf{Score}\,(\%) \\
    \hline
    \multirow{3}{*}{\multirow{3}{*}\textbf{EfficientNet}} & $\text{CE}$ & \multirow{3}{*}{\multirow{3}{*}\textbf{IN}} & $\text{73.48}_{\pm 5.93}$ & $\text{\bf 39.24}_{\pm 2.43}$ & $\text{{57.46}}_{\pm 1.05}$ \\

    & $\text{DAT}$ & & $\text{\bf 89.99}_{\pm 7.15}$ & $\text{11.78}_{\pm 7.05}$ & $\text{50.89}_{\pm 0.69}$ \\

    & $\text{SG-SCL}$ & & $\text{{81.58}}_{\pm 3.38}$ & $\text{{33.85}}_{\pm 3.75}$ & $\text{\bf 57.72}_{\pm 1.32}$ \\
    \bottomrule

    \multirow{3}{*}{\multirow{3}{*}\textbf{ResNet18}} & $\text{CE}$ & \multirow{3}{*}{\multirow{3}{*}\textbf{IN}} & $\text{74.72}_{\pm 3.43}$ & $\text{{33.95}}_{\pm 3.88}$ & $\text{{54.33}}_{\pm 0.91}$ \\

    & $\text{DAT}$ & & $\text{\bf 91.26}_{\pm 6.32}$ & $\text{12.03}_{\pm 5.86}$ & $\text{51.51}_{\pm 0.34}$ \\

    & $\text{SG-SCL}$ & & $\text{{75.58}}_{\pm 6.36}$ & $\text{\bf 34.63}_{\pm 5.98}$ & $\text{\bf 55.10}_{\pm 1.18}$ \\
    \bottomrule

    \multirow{3}{*}{\multirow{3}{*}\textbf{CNN6}} & $\text{CE}$ & \multirow{3}{*}{\multirow{3}{*}\textbf{IN}} & $\text{{80.13}}_{\pm 2.64}$ & $\text{{35.91}}_{\pm 3.52}$ & $\text{{58.10}}_{\pm 0.59}$ \\

    & $\text{DAT}$ & & $\text{\bf 88.57}_{\pm 5.66}$ & $\text{13.73}_{\pm 6.47}$ & $\text{51.15}_{\pm 0.45}$ \\

    & $\text{SG-SCL}$ & & $\text{78.16}_{\pm 3.49}$ & $\text{\bf 38.05}_{\pm 4.41}$ & $\text{\bf 58.11}_{\pm 0.64}$ \\
    \bottomrule

    \multirow{3}{*}{\multirow{3}{*}\textbf{AST}} & $\text{CE}$ & \multirow{3}{*}{\multirow{3}{*}\textbf{\!IN\,+\,AS\!}} & $\text{{77.14}}_{\pm 3.35}$ & $\text{{41.97}}_{\pm 2.21}$ & $\text{59.55}_{\pm 0.88}$ \\

    & $\text{DAT}$ & & $\text{77.11}_{\pm 7.20}$ & $\text{{41.99}}_{\pm 5.00}$ & $\text{{59.81}}_{\pm 1.25}$ \\

    & $\text{SG-SCL}$ & & $\text{\bf 79.87}_{\pm 8.89}$ & $\text{{\bf 43.55}}_{\pm 5.93}$ & $\text{\bf 61.71}_{\pm 1.61}$ \\
    \bottomrule

    
    \end{tabular}
    }
    \vspace{-5mm}
\end{table}


Our experimental results in Table~\ref{tab:target_representations} show that the most promising results were obtained when the stop-gradient was applied to the target representations, $z_{p}$, which are the second augmented samples from the same source sample in the multi-view batch. We found that allowing gradient flows through both anchor and target representations, $z_{i}$ and $z_{p}$, simultaneously did not improve the performance. This suggests that the stop-gradient applied to the target representations $z_{p}$ is more effective in the latent space for cross-domain adaptation.


\vspace{-2mm}
\section{Results}
\vspace{-1mm}
\label{sec:results}

\begin{table*}[!t]
    \centering
    \vspace{-3mm}
    \caption{Comprehensive comparison of the ICBHI dataset for the respiratory sound classification task. We compared previous studies that adhered to the official split of the ICBHI dataset (60-40\% split for the train/test set). Scores marked with $*$ denote the previous state-of-the-art performance. \textbf{Best} and {\underline{second best}} results.}
    
    \label{tab:overall_results}
    \renewcommand{\arraystretch}{1}
    \addtolength{\tabcolsep}{3pt}
    \resizebox{\linewidth}{!}{
    \begin{tabular}{p{2pt}llcllll}
    \toprule
    & method & architecture & pretrain & venue & $S_p$\,(\%) & $S_e$\,(\%) & \textbf{Score}\,(\%) \\
    \hline \midrule
    \multirow{12.5}{*}{\rotatebox[origin=c]{90}{\textbf{4-class eval.}}} & CNN-MoE \cite{pham2021cnn} & C-DNN & - & \textit{JBHI`21} & 72.40 & 21.50 & 47.00 \\
    & RespireNet \cite{gairola2021respirenet} (CBA+BRC+FT) & ResNet34 & IN & \textit{EMBC`21} & 72.30 & 40.10  & 56.20 \\
    & Ren \textit{et al.} \cite{ren2022prototype} & CNN8-Pt & - & \textit{ICASSP`22} & 72.96 & 27.78 & 50.37 \\
    & Chang \textit{et al.} \cite{chang22h_interspeech} & CNN8-dilated & - & \textit{INTERSPEECH`22} & 69.92 & 35.85 & 52.89 \\
    & Chang \textit{et al.} \cite{chang22h_interspeech} & ResNet-dilated & - & \textit{INTERSPEECH`22} & 50.22 & 51.83 & 51.02 \\
    & Wang \textit{et al.} \cite{wang2022domain} (Splice) & ResNeSt & IN & \textit{ICASSP`22} & 70.40 & 40.20 & 55.30 \\
    & Nguyen \textit{et al.} \cite{nguyen2022lung}\,(CoTuning) & ResNet50 & IN & \textit{TBME`22} & 79.34 & 37.24 & $\text{58.29}$ \\
    & Moummad \textit{et al.} \cite{moummad2022supervised}\,(SCL) & CNN6 & AS & \textit{arXiv`22} & 75.95 & 39.15 & 57.55 \\    
    & Bae \textit{et al.} \cite{bae23b_interspeech}\, (Fine-tuning)  & AST & IN\,+\,AS & \textit{INTERSPEECH`23} & $\text{77.14}$ & $\text{41.97}$ & $\text{59.55}$ \\
    & Bae \textit{et al.} \cite{bae23b_interspeech}\, (Patch-Mix CL) & AST & IN\,+\,AS & \textit{INTERSPEECH`23} & $\text{\bf {81.66}}$ & $\text{\underline{43.07}}$ & $\text{\bf 62.37}^\textbf{*}$ \\
    \cmidrule{2-8}
    
    & \textbf{DAT [ours]} & AST & IN\,+\,AS & \textit{ICASSP`24} & $\text{77.11}_{\pm 7.2}$ & $\text{42.50}_{\pm 5.39}$ & $\text{59.81}_{\pm 1.25}$ \\
    & \textbf{SG-SCL [ours]} & AST & IN\,+\,AS & \textit{ICASSP`24} & $\text{\underline{79.87}}_{\pm 8.89}$ & $\text{\bf{43.55}}_{\pm 5.93}$ & $\text{\underline{61.71}}_{\pm 1.61}$ \\

    \midrule
    
    \multirow{6.5}{*}{\rotatebox[origin=c]{90}{\textbf{2-class eval.}}} & 
    CNN-MoE \cite{pham2021cnn} & C-DNN & - & \textit{JBHI`21} & 72.40 & 37.50 & 54.10 \\
    & Nguyen \textit{et al.} \cite{nguyen2022lung}\,(CoTuning) & ResNet50 & IN & \textit{TBME`22} & 79.34 & 50.14 & $\text{64.74}$ \\
    & Bae \textit{et al.} \cite{bae23b_interspeech}\, (Fine-tuning) & AST & IN\,+\,AS & \textit{INTERSPEECH`23} & $\text{77.14}$ & $\text{56.40}$ & $\text{66.77}$ \\
    & Bae \textit{et al.} \cite{bae23b_interspeech}\, (Patch-Mix CL) & AST & IN\,+\,AS & \textit{INTERSPEECH`23} & $\text{\bf 81.66}$ & $\text{55.77}$ & $\text{\underline{68.71}}^\textbf{*}$ \\
    \cmidrule{2-8}
    & \textbf{DAT [ours]} & AST & IN\,+\,AS & \textit{ICASSP`24} & $\text{77.11}_{\pm 7.2}$ & $\text{\underline{56.98}}_{\pm 7.42}$ & $\text{67.04}_{\pm 1.29}$ \\
    & \textbf{SG-SCL [ours]} & AST & IN\,+\,AS & \textit{ICASSP`24} & $\text{\underline{79.87}}_{\pm 8.89}$ & $\text{\bf 57.97}_{\pm 8.96}$ & $\text{\bf 68.93}_{\pm 1.47}$ \\
    \bottomrule
    \end{tabular}}
    \vspace{-5mm}
\end{table*}

\subsection{Effectiveness of Proposed SG-SCL}
To validate the effectiveness of the proposed methods, we trained CE, DAT, and SG-SCL methods on different architectures with the ICBHI dataset under the same conditions without additional learning techniques. As shown in Table~\ref{tab:comparison1}, the proposed SG-SCL method achieved the best Score in all architectures, while the DAT method showed the lowest performance except for the AST architecture. We conjecture that these outcomes are due to potential underfitting issues inherent to the DAT domain classifier, which may come out from the limitation of the CNN-based architectural designs. 

\subsection{Overall ICBHI Dataset Results}
As summarized in Table~\ref{tab:overall_results}, we conducted an overall comparison of the ICBHI dataset for the lung sound classification task, including our proposed methods. In 4-class evaluation, the proposed SG-SCL achieved a 61.71\% Score, which outperforms the AST fine-tuning model~\cite{bae23b_interspeech} by 2.16\%. Interestingly, our SG-SCL achieved a state-of-the-art Score in 2-class evaluation, with a value of 68.93\%. Moreover, our SG-SCL obtained the highest Sensitivity $S_e$ in both 4-class and 2-class evaluations, with the values of 43.55\%, 57.97\%. These results suggest that our SG-SCL is the most accurate model for actually classifying abnormal lung sounds. We would like to highlight that the proposed stethoscope-guided supervised contrastive learning for cross-domain adaptation methodology achieved a significant improvement of 2.16\% over the AST baseline, without any additional learning techniques such as device-specific fine-tuning~\cite{gairola2021respirenet}, co-tuning~\cite{nguyen2022lung}, or Patch-Mix augmentation~\cite{bae23b_interspeech}.

\vspace{-3mm}
\subsection{Qualitative Analysis}
\begin{figure}[!t]
    \vspace{-1mm}
    \centering
    \begin{subfigure}{.5\linewidth}
      \centering
      \includegraphics[width=1.0\linewidth]{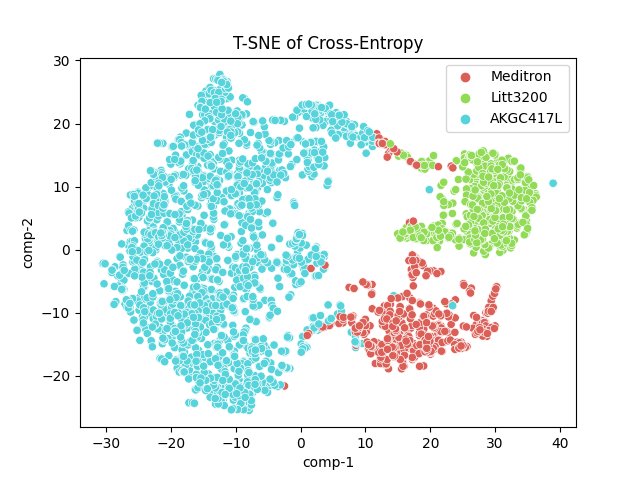}
      \caption{AST fine-tuning}
      \label{fig:sfig1}
    \end{subfigure}%
    \begin{subfigure}{.5\linewidth}
      \centering
      \includegraphics[width=1.0\linewidth]{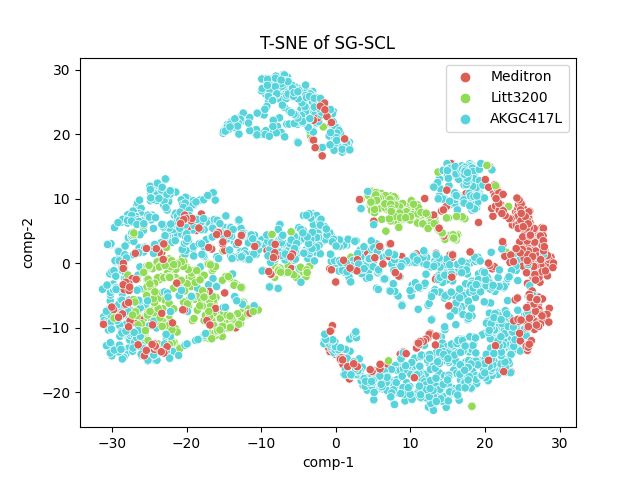}
      \caption{Proposed SG-SCL}
      \label{fig:sfig2}
    \end{subfigure}    
    \vspace{-5mm}
    \caption{T-SNE results of AST fine-tuning and SG-SCL on ICBHI test set for stethoscope labels.}
    \label{fig:t-sne}
    \vspace{-5mm}
    \end{figure}

\vspace{-1mm}
To analyze the extent to which our SG-SCL mitigates domain dependencies, Fig.~\ref{fig:t-sne} provides the t-SNE results generated by the outputs of the feature extractor in terms of stethoscope labels from the ICBHI test set for both the AST fine-tuning and SG-SCL. The AST fine-tuning results in Fig.~\ref{fig:sfig1} show that the representations are clustered according to the stethoscopes, while our SG-SCL results in Fig.~\ref{fig:sfig2} are well mixed regardless of the stethoscope type.

\vspace{-1mm}
\section{Conclusion}
The scarcity of data limits the learning ability of neural networks, and potential biases from various electronic devices for obtaining medical data cause domain dependencies, which can lead to overfitting and poor generalization performance. 
In this paper, we addressed these challenges by introducing domain adversarial training techniques and proposing a novel stethoscope-guided supervised contrastive learning approach for cross-domain adaptation.
Experimental results on the ICBHI dataset demonstrated the effectiveness of the proposed methods, achieving the ICBHI Score of 61.71\%, which is a significant improvement of 2.16\% over the baseline. 
Consequently, this suggested that our method can effectively alleviate domain-related mismatches by learning to distinguish respiratory sounds regardless of the recording devices. 

\vfill\pagebreak

\bibliographystyle{IEEEbib}
\bibliography{strings,refs}

\begin{thebibliography}{10}

\bibitem{rennoll2020electronic}
Valerie Rennoll, Ian McLane, Dimitra Emmanouilidou, James West, and Mounya
  Elhilali,
\newblock ``Electronic stethoscope filtering mimics the perceived sound
  characteristics of acoustic stethoscope,''
\newblock {\em IEEE journal of biomedical and health informatics}, vol. 25, no.
  5, pp. 1542--1549, 2020.

\bibitem{seah2023review}
Jun~Jie Seah, Jiale Zhao, De~Yun Wang, and Heow~Pueh Lee,
\newblock ``Review on the advancements of stethoscope types in chest
  auscultation,''
\newblock {\em Diagnostics}, vol. 13, no. 9, pp. 1545, 2023.

\bibitem{gairola2021respirenet}
Siddhartha Gairola, Francis Tom, Nipun Kwatra, and Mohit Jain,
\newblock ``Respirenet: A deep neural network for accurately detecting abnormal
  lung sounds in limited data setting,''
\newblock in {\em 2021 43rd Annual International Conference of the IEEE
  Engineering in Medicine \& Biology Society (EMBC)}. IEEE, 2021, pp. 527--530.

\bibitem{ren2022prototype}
Zhao Ren, Thanh~Tam Nguyen, and Wolfgang Nejdl,
\newblock ``Prototype learning for interpretable respiratory sound analysis,''
\newblock in {\em ICASSP 2022-2022 IEEE International Conference on Acoustics,
  Speech and Signal Processing (ICASSP)}. IEEE, 2022, pp. 9087--9091.

\bibitem{wang2022domain}
Zijie Wang and Zhao Wang,
\newblock ``A domain transfer based data augmentation method for automated
  respiratory classification,''
\newblock in {\em ICASSP 2022-2022 IEEE International Conference on Acoustics,
  Speech and Signal Processing (ICASSP)}. IEEE, 2022, pp. 9017--9021.

\bibitem{nguyen2022lung}
Truc Nguyen and Franz Pernkopf,
\newblock ``Lung sound classification using co-tuning and stochastic
  normalization,''
\newblock {\em IEEE Transactions on Biomedical Engineering}, vol. 69, no. 9,
  pp. 2872--2882, 2022.

\bibitem{bae23b_interspeech}
Sangmin Bae, June-Woo Kim, Won-Yang Cho, Hyerim Baek, Soyoun Son, Byungjo Lee,
  Changwan Ha, Kyongpil Tae, Sungnyun Kim, and Se-Young Yun,
\newblock ``{Patch-Mix Contrastive Learning with Audio Spectrogram Transformer
  on Respiratory Sound Classification},''
\newblock in {\em Proc. INTERSPEECH 2023}, 2023, pp. 5436--5440.

\bibitem{guan2021domain}
Hao Guan and Mingxia Liu,
\newblock ``Domain adaptation for medical image analysis: a survey,''
\newblock {\em IEEE Transactions on Biomedical Engineering}, vol. 69, no. 3,
  pp. 1173--1185, 2021.

\bibitem{quinonero2008dataset}
Joaquin Quinonero-Candela, Masashi Sugiyama, Anton Schwaighofer, and Neil~D
  Lawrence,
\newblock {\em Dataset shift in machine learning},
\newblock Mit Press, 2008.

\bibitem{torralba2011unbiased}
Antonio Torralba and Alexei~A Efros,
\newblock ``Unbiased look at dataset bias,''
\newblock in {\em CVPR 2011}. IEEE, 2011, pp. 1521--1528.

\bibitem{ganin2016domain}
Yaroslav Ganin, Evgeniya Ustinova, Hana Ajakan, Pascal Germain, Hugo
  Larochelle, Fran{\c{c}}ois Laviolette, Mario Marchand, and Victor Lempitsky,
\newblock ``Domain-adversarial training of neural networks,''
\newblock {\em The journal of machine learning research}, vol. 17, no. 1, pp.
  2096--2030, 2016.

\bibitem{khosla2020supervised}
Prannay Khosla, Piotr Teterwak, Chen Wang, Aaron Sarna, Yonglong Tian, Phillip
  Isola, Aaron Maschinot, Ce~Liu, and Dilip Krishnan,
\newblock ``Supervised contrastive learning,''
\newblock {\em Advances in neural information processing systems}, vol. 33, pp.
  18661--18673, 2020.

\bibitem{rocha2018alpha}
BM~Rocha, Dimitris Filos, L~Mendes, Ioannis Vogiatzis, Eleni Perantoni,
  E~Kaimakamis, P~Natsiavas, Ana Oliveira, C~J{\'a}come, A~Marques, et~al.,
\newblock ``A respiratory sound database for the development of automated
  classification,''
\newblock in {\em Precision Medicine Powered by pHealth and Connected Health:
  ICBHI 2017, Thessaloniki, Greece, 18-21 November 2017}. Springer, 2018, pp.
  33--37.

\bibitem{gong21b_interspeech}
Yuan Gong, Yu-An Chung, and James Glass,
\newblock ``{AST: Audio Spectrogram Transformer},''
\newblock in {\em Proc. Interspeech 2021}, 2021, pp. 571--575.

\bibitem{park2019specaugment}
Daniel~S. Park, William Chan, Yu~Zhang, Chung-Cheng Chiu, Barret Zoph, Ekin~D.
  Cubuk, and Quoc~V. Le,
\newblock ``Specaugment: A simple data augmentation method for automatic speech
  recognition,''
\newblock {\em Interspeech 2019}, 2019.

\bibitem{moummad2022supervised}
Ilyass Moummad and Nicolas Farrugia,
\newblock ``Supervised contrastive learning for respiratory sound
  classification,''
\newblock {\em arXiv preprint arXiv:2210.16192}, 2022.

\bibitem{deng2009imagenet}
Jia Deng, Wei Dong, Richard Socher, Li-Jia Li, Kai Li, and Li~Fei-Fei,
\newblock ``Imagenet: A large-scale hierarchical image database,''
\newblock in {\em 2009 IEEE conference on computer vision and pattern
  recognition}. Ieee, 2009, pp. 248--255.

\bibitem{audioset}
Jort~F. Gemmeke, Daniel P.~W. Ellis, Dylan Freedman, Aren Jansen, Wade
  Lawrence, R.~Channing Moore, Manoj Plakal, and Marvin Ritter,
\newblock ``Audio set: An ontology and human-labeled dataset for audio
  events,''
\newblock in {\em Proc. IEEE ICASSP 2017}, New Orleans, LA, 2017.

\bibitem{pham2021cnn}
Lam Pham, Huy Phan, Ramaswamy Palaniappan, Alfred Mertins, and Ian McLoughlin,
\newblock ``Cnn-moe based framework for classification of respiratory anomalies
  and lung disease detection,''
\newblock {\em IEEE journal of biomedical and health informatics}, vol. 25, no.
  8, pp. 2938--2947, 2021.

\bibitem{chang22h_interspeech}
Yi~Chang, Zhao Ren, Thanh~Tam Nguyen, Wolfgang Nejdl, and Björn~W. Schuller,
\newblock ``{Example-based Explanations with Adversarial Attacks for
  Respiratory Sound Analysis},''
\newblock in {\em Proc. Interspeech 2022}, 2022, pp. 4003--4007.

\end{thebibliography}

\end{document}